\documentclass[12pt]{article}
\usepackage[dvipdfmx]{graphicx}
\begin{document}
\title{A General Derivation of Pointer States:\\
Decoherence and Classicality} 
\author{{Kentaro Urasaki}%
\footnote{Yokohama National University, urasaki@ynu.ac.jp}}
\date{}%

\maketitle

\abstract{
The purpose of the present study is to derive the pointer states of a macroscopic system interacting with its environment, under the general assumptions, i.e., without assuming any form of the interaction Hamiltonian. 
The lowest order perturbation leads that the interaction energy shifts the phase factors of the state vectors.  
For a macroscopic system, these factors are the macroscopic quantities even for the very weak interaction. 
When we group the state vector of the total system by the view point of environmental side, the destructive interference occurs and the stationary phase approximation can be adopted. 
Only the pointer states then survive and the decoherence also occurs.
The present approach is within the standard quantum mechanics as same as the standard decoherence theory, but the meaning of the classical state is much clear.
}

\section{Pointer states and classicality}

The problem of pointer states is much relevant  to the quantum origin of the classical world.
Among very many degrees of freedom of a macroscopic system,
small number of them behave classical.
The classical states of these degrees of freedom are called as pointer states.
A typical example is the center of mass position of a macroscopic object.
It does not behave as quantum mechanically, 
superposition of states, but has a certain position.
On the other hands, other almost all of the degrees of freedom are true to quantum mechanics.

It has already been pointed out that the eigenstates of the interaction Hamiltonian between a macro-system and its environment play important role: 
On this subject, 
after the 1970s the idea of decoherence have been studied\cite{Joos1985}\cite{Zurek1981} to understand 
the classical features of our daily experiences within quantum mechanics. 
Since the decoherence theory\cite{Joos1985}\cite{Zurek1981} is conceptually close to our approach,
we shortly refer this below.

\subsection{Pointer states and decoherence theory}

For the two states of the macroscopic system, 
we consider the interaction with its environment.
The initial states of the total system is, 
\begin{eqnarray}
|\Phi(0)\rangle=(c_1|\phi_1(0)\rangle+c_2|\phi_2(0)\rangle)|\varepsilon(0)\rangle.
\end{eqnarray}
If we assume an appropriate interaction, 
it evolves into 
\begin{eqnarray}\label{eq:vNform}
|\Phi(t)\rangle=c_1|\phi_1(t)\rangle|\varepsilon_1(t)\rangle+c_2|\phi_2(t)\rangle
|\varepsilon_2(t)\rangle.
\end{eqnarray}
Although this standard expression (von Neumann form) focusing the states of the macroscopic system is often used, in the next section, we express the same equation in another form focusing the state of the environment.

We notice that these two states are assumed to be stable against the interaction.
Therefore it is important that these states, $|\phi_1\rangle, |\phi_2\rangle$, in generally, may be  approximate eigenstates of the interaction Hamiltonian.
These states are called as the pointer states and essential role in the decoherence scenario.

For the operator $\hat{Q}$ acting only on the subsystem $\phi$, 
its expectation value is expressed as, 
\begin{eqnarray}
\langle Q \rangle=|c_1|^2\langle\phi_1|\hat{Q}|\phi_1\rangle+|c_2|^2\langle\phi_2|\hat{Q}|\phi_2\rangle
+c^\ast_1 c_2\langle\phi_1|\hat{Q}|\phi_2\rangle\langle\varepsilon_1|\varepsilon_2\rangle
+c^\ast_2 c_1\langle\phi_2|\hat{Q}|\phi_1\rangle\langle\varepsilon_2|\varepsilon_1\rangle.
\end{eqnarray}
If the state of the environment develop into the corresponding orthogonal state
after the interaction, we can say, 
\begin{eqnarray}
\langle\varepsilon_1|\varepsilon_2\rangle\sim 0.
\end{eqnarray}
In this case, the coherent terms in the above equation become small. 
This process is called decoherence.
Since in this case the above equation is similar to that of statistical mixture of the events, we may also say the approximate mixture is obtained. 
We also find out the same result  
introducing the reduced density operator for the macroscopic system as,
\begin{eqnarray}
\rho_\phi:={\rm Tr}_\varepsilon\rho_\Phi,
\end{eqnarray}
where the density matrix for mixed states appears. 

The decoherence theory is studied intensively and approved by no small number of researchers because of its conceptual simplicity, where the time evolution is under the total system-environment Hamiltonian. 
The many facts have been clarified\cite{Joos1985}\cite{Zurek1981}
and its relation to the interpretation of quantum mechanics is also discussed\cite{Schlosshauer2005}.
On the other hand, there are still some essential questions.
For example, in quantum mechanics, in principle, any choice of basis give same results.
Namely, if  nothing happens in certain basis, we can conclude nothing happens.
In the decoherence theory, however, only the pointer states seem to be effective and capable to describe our reality. 
Why do we experience specific basis? 
The mechanism of the emergence of the pointer states still seems unclear
to be suitable for textbooks.

Although we agree with the basic idea of the decoherence theory,  
we examine the starting point below, where we find that 
one experiences the pointer states not because of its effectiveness but because of only the pointer states surviving.

\section{Formulation}
\subsection{(already entangled) Initial state}
Assuming that a macroscopic system is described by two orthogonal states and the environmental system is described by $N$ orthogonal states.
(We can easily generalize the number of states of the macroscopic system.)
Each of the $2N$ eigenstates of the interaction Hamiltonian is decomposed into the product states as, 
$|\uparrow\rangle|\varepsilon_1\rangle, \cdots, |\downarrow\rangle|\varepsilon_N\rangle,$
where $|\uparrow\rangle, |\downarrow\rangle$
correspond to the states of the macroscopic system and 
$|\varepsilon_1\rangle, \cdots, |\varepsilon_N\rangle$ correspond to the states of the environment.

Using these states, we start from the initial state,
\begin{eqnarray}\label{eq:entangled}
|\Phi(0)\rangle=C_{\uparrow 1}|\uparrow\rangle|\varepsilon_1\rangle+\cdots+
C_{\downarrow N}|\downarrow\rangle|\varepsilon_N\rangle,
\end{eqnarray}
representing general entangled states.
Thia is reasonable if we remember that the macroscopic system interacts with the environment continuously.  
Generally any state of center of mass interacted complicatedly with other microscopic states in the past.

Next, we separate this to $N$ states from the viewpoint of the environment side, 
\begin{eqnarray}
|\Phi(0)\rangle&&=(C_{\uparrow 1}|\uparrow\rangle+C_{\downarrow 1}|\downarrow\rangle)
|\varepsilon_1\rangle
+\cdots+(C_{\uparrow N}|\uparrow\rangle+C_{\downarrow N}|\downarrow\rangle)
|\varepsilon_N\rangle\\
&&=:\sum_{\nu=1}^N\alpha_\nu|\nu\rangle, 
\end{eqnarray}
where the states $|\nu\rangle=(c_{\uparrow\nu}|\uparrow\rangle+c_{\downarrow\nu}|\downarrow\rangle)|\varepsilon_\nu\rangle$ are appropriately normalized and orthogonal each other. 
It is important that {\bf various superposition states of the macroscopic system} appear in this expression. 

This is, in a sense, the inverse version of eq. (\ref{eq:vNform}):
Namely, in usual discussions, such correlated state is often represented as, 
\begin{eqnarray}
=|\uparrow\rangle(C_{\uparrow 1}|\varepsilon_1\rangle+\cdots+C_{\uparrow N}| \varepsilon_N\rangle)+|\downarrow\rangle(C_{\downarrow 1}|\varepsilon_1\rangle+\cdots +C_{\downarrow N}|\varepsilon_N\rangle), 
\end{eqnarray}
being grouped into two groups.

\vspace{1cm}
\underline{\bf Notice}\\
For comparison, if we start from the product state, which is a very special case, we obtain 
\begin{eqnarray}
|\phi\rangle\otimes|\varepsilon\rangle
&&
=(c_\uparrow|\uparrow\rangle+c_\downarrow|\downarrow\rangle)(c_1|\varepsilon_1\rangle+ \cdots+c_N|\varepsilon_N\rangle)\\
&&
=\sum_{\nu=1}^Nc_\nu(c_\uparrow|\uparrow\rangle+c_\downarrow|\downarrow\rangle)| \varepsilon_\nu\rangle.
\end{eqnarray}
Here we see that each state corresponds same superposition state of the macroscopic system.
This, however, is wrong at least with respect to the center of mass position as mentioned above. 
We can show the emergence of the classical state, paradoxically starting from the sufficiently 
entangled state eq. (\ref{eq:entangled}) below.

\subsection{Interaction with the environment}
The total system obeys the Schr\"odinger equation,
\begin{eqnarray}
[i\hbar\partial_t-(\hat{h}_\phi+\hat{h}_{int}+\hat{h}_\varepsilon)]|\Phi(t)\rangle=0.
\end{eqnarray}
We expand $|\Phi(t)\rangle$ by the non-perturbative states, 
\begin{eqnarray}
|\nu(t)\rangle=(c_{\uparrow\nu}|\uparrow(t)\rangle+c_{\downarrow\nu}|\downarrow (t)\rangle)|\varepsilon_\nu(t)\rangle,
\end{eqnarray}
assumed to be orthogonal to each other.\footnote{At $t=0$, this holds exactly.} 
Substituting $|\Phi(t)\rangle=\sum_{\nu=1}^N\alpha_\nu(t)|\nu(t)\rangle$
and acting $\langle\nu(t)|$ from the left side, we obtain 
\begin{eqnarray}
i\partial_t \alpha_\nu(t)=\sum_{\nu^\prime}\alpha_{\nu^\prime}(t)\langle \nu(t)|\hat{h}_{int}|\nu^\prime(t)\rangle\ \ \ \simeq \ \  \alpha_\nu(t)\langle \nu(t)|\hat{h}_{int}|\nu(t)\rangle.
\end{eqnarray}
Although we treat the interaction effect perturbatively, 
in the last equation, we neglect transition,\footnote{At $t=0$, this also holds exactly.} 
\begin{eqnarray}
\langle\nu(t)|\hat{h}_{int}|\nu^\prime(t)\rangle\sim 0.
\end{eqnarray}
We can very easily integrate this and obtain
$\displaystyle \alpha_\nu(t)=\alpha_{\nu}\exp\left[-i\int \langle \nu(t)|\hat{h}_{int}|\nu(t)\rangle dt/\hbar\right].$
This treatment is equivalent to the lowest order approach in the perturbation theory, where only the phase shift due to the interaction energy is taken into account.
After all,
\begin{eqnarray}
|\Phi(t)\rangle\simeq\sum_{\nu=1}^N\alpha_\nu|\nu(t)\rangle e^{-i\Lambda_\nu(t)/\hbar}, 
\end{eqnarray}
where $\Lambda_\nu(t):=\int\langle\nu(t)|\hat{h}_{int}|\nu(t)\rangle dt$
represents the time integral of the interaction energy. 
Even for the very weak interaction, $\Lambda_\nu(t)$ is the macroscopic quantity and 
occurs very frequent sign inversion.

\subsection{Destructive interference due to the phase shift}
Therefore only the states giving the extreme values to $\Lambda_\nu(t)$ can survive. 
The stationary phase approximation leads, 
\begin{eqnarray}
|\Phi(t)\rangle\simeq\sum_{\nu_c}\tilde{\alpha}_{\nu_c}|\nu_c(t)\rangle e^{-i\Lambda_{\nu_c}(t)/\hbar}.
\end{eqnarray}
Finally almost all of the states vanish except for the classical states $|\nu_c(t)\rangle$.
(It is similar to the case that the classical limit in the path integral formulation.)
These states are the approximate eigenstates of the interaction Hamiltonian, so-called pointer states. 
This result is obviously independent of the representation.
In this point our result disparate to the standard decoherence scenario.

For simplicity, we assume that the interaction energy depends only on 
the state of the macroscopic system, $|\phi(t)\rangle$.
In this case, introducing the external potential, 
\begin{eqnarray}
\hat{V}:=\langle\varepsilon_\nu|\hat{h}_{int}|\varepsilon_\nu\rangle, 
\end{eqnarray}
\begin{eqnarray}
\Lambda_\nu(t)=\int(|c_{\uparrow\nu}|^2\langle\uparrow(t)|\hat{V}|\uparrow(t)\rangle+|c_ {\downarrow\nu}|^2\langle\downarrow(t)|\hat{V}|\downarrow(t)\rangle)dt.
\end{eqnarray}
Except for the case of accidental degeneracy in the interaction enegy, 
the extreme values occur at $c_{\uparrow\nu}=0$ or $c_ {\downarrow\nu}=0$ and then,
\begin{eqnarray}\label{eq:nuc}
|\Phi(t)\rangle\simeq\sum_{c_{\downarrow\nu}=0}\tilde\alpha_{\nu_c}|\uparrow(t)\rangle|\varepsilon_{\nu_c}(t)\rangle e^{-i\Lambda_{\nu_c}(t)/\hbar}
+\sum_{c_{\uparrow\nu}=0}\tilde\alpha_{\nu_c}|\downarrow(t)\rangle|\varepsilon_{\nu_c}(t)\rangle e^{-i\Lambda_{\nu_c}(t)/\hbar}.
\end{eqnarray}
Surviving states are classical as seen in Fig. 1. 
\footnote{Although  the coefficients $\{c_{\uparrow\nu}, c_ {\downarrow\nu}\}$ are unknown, we here exclude the exceptional distributions. }

\begin{figure}
\begin{center}
\includegraphics[width=7cm]{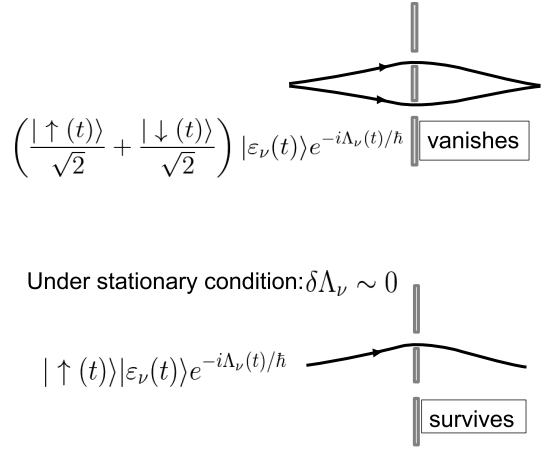}
\caption{The contribution of the stationary phase (image).}
\end{center}
\end{figure}

\vspace{1cm}
\underline{\bf Localization of the center of mass}\\
Although we have studied on the two-state macroscopic system above, 
the obtained results are easily extended to the continuous case.
Representing the center of mass position with the wave function, 
the time-integrated  interaction energy is, 
\begin{eqnarray}
\Lambda_\nu(t)=\int\int\phi^\ast({\bf R}, t)V({\bf R})\phi({\bf R}, t)d{\bf R}dt
\to|\phi({\bf R})|^2\sim\delta({\bf R}-{\bf R}_c).
\end{eqnarray}
In this case, however, the delocalization by the non-pertubative Hamiltonian and the localization by the destructive interference are competitive.

\subsection{Decoherence in the present formulation}

For the comparison purposes, we can re-order the terms in eq. (\ref{eq:nuc})  to correspond to Schmidt representation\footnote{In our original formulation, each term, $|\nu_c(t)\rangle e^{-i\Lambda_{\nu_c}(t)/\hbar}$, represents a classical state.}, 
\begin{eqnarray}
|\Phi(t)\rangle
&&\simeq\sum_{c_{\downarrow\nu}=0}\tilde\alpha_{\nu_c}|\uparrow(t)\rangle|\varepsilon_{\nu_c}(t)\rangle e^{-i\Lambda_{\nu_c}(t)/\hbar}
+\sum_{c_{\uparrow\nu}=0}\tilde\alpha_{\nu_c}|\downarrow(t)\rangle|\varepsilon_{\nu_c}(t)\rangle e^{-i\Lambda_{\nu_c}(t)/\hbar}\\
&&=:\tilde\alpha_{\uparrow}|\uparrow(t)\rangle|\varepsilon_A(t)\rangle+\tilde\alpha_{\downarrow}|\downarrow(t)\rangle|\varepsilon_B(t)\rangle, 
\end{eqnarray}
where $|\varepsilon_A(t)\rangle:=\sum_{c_{\uparrow\nu}=0}\tilde\alpha_{\nu_c}|\varepsilon_{\nu_c}(t)\rangle e^{-i\Lambda_{\nu_c}(t)/\hbar}$ and $|\varepsilon_B(t)\rangle:=\sum_{c_{\uparrow\nu}=0}\tilde\alpha_{\nu_c}|\varepsilon_{\nu_c}(t)\rangle e^{-i\Lambda_{\nu_c}(t)/\hbar}$.
These states are paradoxically stabilized by the interaction. 

Since the states separated to $|\varepsilon_A\rangle$ and $|\varepsilon_B\rangle$ have already vanished by the destructive interference, 
 $\langle\varepsilon_B|\varepsilon_A\rangle=0$ and 
$\langle\varepsilon_B(t)|\varepsilon_A(t)\rangle\simeq 0$
is lead.
The present study clearly reproduce the results of the previous studies in decoherence\cite{Joos1985}\cite{Zurek1981}, giving the precise meaning of the classical states for general case.

\section{Conclusions}
Under the general but weak interaction with the environment, we have illustrated the process of the emergence of the pointer states taking into account of the time-integrated interaction energy $\Lambda_\nu(t)$, where only the state giving extreme values to $\Lambda_\nu(t)$ survive.
The classical states are clearly identified.

\vspace{1cm}

Other main conclusions are, 
\begin{itemize}
\item{Like the center of mass position, the states almost isolated but weakly interacting with its environment behave classical.}
\item{The decoherence is also derived.}
\item{Our results are independent of the representation (basis).}
\end{itemize}

\vspace{1cm}

On the other hand, the present study is based on the following essential assumptions: 
\begin{itemize}
\item{The time evolution under the Schr\"odinger equation}
\item{The environment-based grouping of the entangled initial state}
\item{The phase shift due to the interaction energy}
\end{itemize}
Especially the meaning of the second assumption will have to be examined.
Note that since the resulting states are in the superposition, 
the interpretation is still left.


\begin{thebibliography}{99}
\bibitem{Joos1985}E. Joos and H. D. Zeh, Z. Phys. B {\bf 59} (1985) 223; 
Joos, E., H. D. Zeh, C. Kiefer, D. Giulini, J. Kupsch, and
I.-O. Stamatescu, 2003, {\it Decoherence and the Appearance
of a Classical World in Quantum Theory} (Springer, New York), 2nd edition.
\bibitem{Zurek1981}W. H. Zurek, Phys. Rev. D {\bf 24} (1981) 1516; W. H. Zurek, Phys. Rev. D {\bf 24} (1982) 1862; W. H. Zurek, Rev. Mod. Phys. {\bf 75} (2003) 715.
\bibitem{Schlosshauer2005} M. Schl\"osshauer, Rev. Mod. Phys. {\bf 76} (2005) 1267.



\end{thebibliography}
\end{document}